\documentclass[11pt,a4paper,showpacs,showkeys,,superscriptaddress]{article}

\usepackage{epsfig}
\usepackage{amsmath,amssymb} 
\usepackage{graphicx}
\usepackage{xcolor}
\usepackage{subfigure}
\usepackage{amstext}
\usepackage{mathrsfs}

\usepackage{latexsym}
\usepackage{hyperref}
\hypersetup{colorlinks,%
  citecolor=blue,%
  linkcolor=red,%
  pdftex}

\begin{document}
\baselineskip 6mm

\newcommand{\nc}{\newcommand}
\newcommand{\rnc}{\renewcommand}



\newcommand{\tcb}{\textcolor{blue}}
\newcommand{\tcr}{\textcolor{red}}
\newcommand{\tcg}{\textcolor{green}}


\def\beq{\begin{equation}}
\def\eeq{\end{equation}}
\def\ba{\begin{array}}
\def\ea{\end{array}}
\def\bea{\begin{eqnarray}}
\def\eea{\end{eqnarray}}
\def\nn{\nonumber}


\def\CMP{Commun. Math. Phys.~}
\def\JHEP{JHEP~}
\def\Pre{Preprint}
\def\PRL{Phys. Rev. Lett.~}
\def\PR {Phys. Rev.~}
\def\CQG {Class. Quant. Grav.~}
\def\PL {Phys. Lett.~}
\def\NP {Nucl. Phys.~}

\def\G{\Gamma}

\def\S{{\bf S}}
\def\C{{\bf C}}
\def\Z{{\bf Z}}
\def\R{{\bf R}}
\def\N{{\bf N}}
\def\M{{\bf M}}
\def\P{{\bf P}}
\def\bm{{\bf m}}
\def\bn{{\bf n}}

\def\CA{{\cal A}}
\def\CB{{\cal B}}
\def\CC{{\cal C}}
\def\CD{{\cal D}}
\def\CE{{\cal E}}
\def\CF{{\cal F}}
\def\CH{{\cal H}}
\def\CM{{\cal M}}
\def\CG{{\cal G}}
\def\CI{{\cal I}}
\def\CJ{{\cal J}}
\def\CL{{\cal L}}
\def\CK{{\cal K}}
\def\CN{{\cal N}}
\def\CO{{\cal O}}
\def\CP{{\cal P}}
\def\CQ{{\cal Q}}
\def\CR{{\cal R}}
\def\CS{{\cal S}}
\def\CT{{\cal T}}
\def\CU{{\cal U}}
\def\CV{{\cal V}}
\def\CW{{\cal W}}
\def\CX{{\cal X}}
\def\CY{{\cal Y}}
\def\CZ{{\cal Z}}

\def\We{{W_{\mbox{eff}}}}


\newcommand{\Lie}{\pounds}

\newcommand{\p}{\partial}
\newcommand{\bp}{\bar{\partial}}

\newcommand{\half}{\frac{1}{2}}

\newcommand{\bfalpha}{{\mbox{\boldmath $\alpha$}}}
\newcommand{\bfbeta}{{\mbox{\boldmath $\beta$}}}
\newcommand{\bfgamma}{{\mbox{\boldmath $\gamma$}}}
\newcommand{\bfmu}{{\mbox{\boldmath $\mu$}}}
\newcommand{\bfpi}{{\mbox{\boldmath $\pi$}}}
\newcommand{\bfvarpi}{{\mbox{\boldmath $\varpi$}}}
\newcommand{\bftau}{{\mbox{\boldmath $\tau$}}}
\newcommand{\bfeta}{{\mbox{\boldmath $\eta$}}}
\newcommand{\bfxi}{{\mbox{\boldmath $\xi$}}}
\newcommand{\bfkappa}{{\mbox{\boldmath $\kappa$}}}
\newcommand{\bfepsilon}{{\mbox{\boldmath $\epsilon$}}}
\newcommand{\bfTheta}{{\mbox{\boldmath $\Theta$}}}

\newcommand{\bz}{{\bar{z}}}

\newcommand{\dalpha}{\dot{\alpha}}
\newcommand{\dbeta}{\dot{\beta}}
\newcommand{\blambda}{\bar{\lambda}}
\newcommand{\btheta}{{\bar{\theta}}}
\newcommand{\bsigma}{{{\bar{\sigma}}}}
\newcommand{\bepsilon}{{\bar{\epsilon}}}
\newcommand{\bpsi}{{\bar{\psi}}}


\def\ct{\cite}
\def\la{\label}
\def\eq#1{(\ref{#1})}


\def\a{\alpha}
\def\b{\beta}
\def\g{\gamma}
\def\G{\Gamma}
\def\d{\delta}
\def\D{\Delta}
\def\ep{\epsilon}
\def\e{\eta}
\def\ph{\phi}
\def\Ph{\Phi}
\def\ps{\psi}
\def\Ps{\Psi}
\def\k{\kappa}
\def\l{\lambda}
\def\L{\Lambda}
\def\m{\mu}
\def\n{\nu}
\def\th{\theta}
\def\Th{\Theta}
\def\r{\rho}
\def\s{\sigma}
\def\S{\Sigma}
\def\ta{\tau}
\def\o{\omega}
\def\O{\Omega}
\def\pr{\prime}
\def\f{\varphi}


\def\half{\frac{1}{2}}

\def\goto{\rightarrow}

\def\na{\nabla}
\def\grad{\nabla}
\def\curl{\nabla\times}
\def\div{\nabla\cdot}
\def\pa{\partial}

\def\bra{\left\langle}
\def\ket{\right\rangle}
\def\lb{\left[}
\def\lc{\left\{}
\def\ls{\left(}
\def\lp{\left.}
\def\rp{\right.}
\def\rb{\right]}
\def\rc{\right\}}
\def\rs{\right)}
\def\cl{\mathcal{l}}

\def\vac#1{\mid #1 \rangle}

\def\td#1{\tilde{#1}}
\def\check{ \maltese {\bf Check!}}


\def\Tr{{\rm Tr}\,}
\def\det{{\rm det}\,}


\def\bc#1{\nnindent {\bf $\bullet$ #1} \\ }
\def\ch {$<Check!>$ }
\def\ss {\vspace{1.5cm}}

\begin{titlepage}
%
%
%
%
%
%
%
%
\begin{center}
{\Large \bf Revisit to Thermodynamic Relations in the AdS/CMT Models}
%
\vskip 1. cm
  {  Seungjoon Hyun\footnote{e-mail : sjhyun@yonsei.ac.kr}, Sang-A Park\footnote{e-mail : sangapark@yonsei.ac.kr},
  Sang-Heon Yi\footnote{e-mail : shyi@yonsei.ac.kr} 
 }
\vskip 0.5cm
{\it Department of Physics, College of Science, Yonsei University, Seoul 120-749, Korea}\\
\end{center}
\thispagestyle{empty}
\vskip1.5cm
%
%
\centerline{\bf ABSTRACT} \vskip 4mm 
 \vspace{1cm} 
\noindent  Motivated by  the recent unified approach to the Smarr-like relation of AdS planar black holes in conjunction with the quasi-local formalism on conserved charges,  we revisit the quantum statistical and thermodynamic relations of hairy AdS planar black holes. By extending the previous results, we identify  the  hairy contribution in the bulk and show that the holographic computation can be  improved  so that it is consistent with the bulk computation.  We argue that the first law can be retained in its universal form, and that the relation between the on-shell renormalized Euclidean action and its free energy interpretation in gravity may also be undeformed even with the hairy contribution in hairy AdS black holes.

\vspace{2cm} 
%
%
\end{titlepage}
\renewcommand{\thefootnote}{\arabic{footnote}}
\setcounter{footnote}{0}
%
%
%
%

\section{Introduction}

It has been known that black holes behave just like thermal objects after Bekenstein and Hawking's pioneering works on their physical properties~\cite{Bekenstein:1973ur,Hawking:1974sw}. This thermodynamic behavior of black holes is regarded as a universal nature of black holes reflecting the hidden physics of black holes.

One of the universal formula in the black hole physics is the expression of the first law of black hole thermodynamics. For Kerr-Newman black holes of the mass $M$, the angular momentum $J$, the $U(1)$ charge $Q$ and  the Bekenstein-Hawking entropy ${\cal S}$, the first law takes the following form 
\[   
dM =T_{H}d{\cal S} +  \Omega dJ +  \mu dQ\,,
\]
where $T_{H}$, $ \Omega$ and $\mu$ denote the Hawking temperature given by the surface gravity, the angular velocity at the horizon and  the $U(1)$ chemical potential.  It would be amusing to  recall that the no-hair theorem of black holes in the asymptotically flat spacetime is consistent with the above form of the first law. 
Based on the black hole uniqueness theorem,  the  first law is established in Einstein gravity~\cite{Bardeen:1973gs} and then it  is shown to hold  in the generic covariant theory of gravity~\cite{Wald:1993nt,Iyer:1994ys}.  Now, it is regarded that the first law in the above form  is the universal property of black holes. This form of the first law has been checked in various black holes even for the asymptotically non-flat spacetime, for instance, for AdS or Lifshitz  black holes.

Another interesting aspect of the thermodynamic relation in black hole physics is the interpretation of the (renormalized) Euclideanized on-shell action value   as the free energy or thermodynamic potential  in (grand) canonical ensemble.  This relation, which is usually called as the {\it quantum statistical relation}, could be written as~\cite{Gibbons:1976ue}
\[   
\frac{1}{\beta}I_{r} = M -T_{H}S - \Omega J  -\mu Q\,,
\]
where $I_{r}$ denotes the on-shell action value and $\beta$ is the periodicity of the Euclideanized time coordinate and is related  to the Hawking temperature as $T_{H}=1/\beta$. Though this relation may be very plausible from the analogy with the path integral formulation of  finite temperature quantum field theories in the flat spacetime, the validity of the relation  is not warranted  partially because the quantum formulation of gravity is not yet accomplished. Nevertheless, one can establish the validity of the relation for Kerr-Newman black holes in the asymptotically flat spacetime.  Moreover, the quantum statistical relation as well as  the first law  are  established rigorously even for the asymptotically AdS rotating charged black holes~\cite{Gibbons:2004ai,Papadimitriou:2005ii}. 
{
On the contrary to the asymptotically flat spacetime, it has been known that hairy black holes exist in the asymptotically AdS spacetime and in fact, these black holes are essential ingredients to various AdS/CMT models. Especially, the holographic superconductors utilize hairy black holes to realize the Cooper pair condensate as the dual to the scalar hair in the AdS black holes. These hairy configurations arouse us the following question: {\it Where are hairy contributions in the above forms of the first law and/or the quantum statistical relation?}  

In this paper, we would like to address this question and show that  various results on the thermodynamic relations in the AdS/CMT models can be interpreted in such a way that the hairy contribution  appears as the deformation in the expression of  conserved charges, but not in the first law nor in the quantum statistical relation. This interpretation turns out to be consistent with the Smarr-like relation  which is shown as a kind of another universal relation among charges on the AdS planar black holes. We will present explicit examples to show that various results on the planar black holes in the AdS/CMT models can be understood in our framework.

\section{Review}

As is well-known~\cite{Gibbons:1976ue}, the path integral approach to quantum mechanics tells us that the Euclidean action  for gravity corresponds to the thermodynamic potential  in the system  under consideration by identifying the periodicity  $\beta$ of the  Euclidean time $\tau$ with the temperature $T=\frac{1}{\beta}$. In the setup of the grand canonical ensemble, one may also  introduce various chemical potentials $\mu_{i}$ associated with  charges $C_{i}$, which are conserved in the sense that $[H,C_{i}]=0$.  The charge, $q^{\Psi}_{i}$, of the field $\Psi$ may be specified as $[C_{i},\Psi] = iq^{\Psi}_{i}\Psi$ in the operator language.  The field $\Psi$ should satisfy the appropriate twisted boundary condition $\Psi(\tau+\beta) = e^{\sum_{i}\mu_{i}q^{\Psi}_{i}}\Psi(\tau)$ along the Euclidean time direction.

Then, the (grand canonical) partition function $Z(\beta,\mu_{i})$  is represented by the path integral as
\[   
Z(\beta,\mu_{i}) = \Tr e^{-\beta(H -\sum_{i}\mu_{i}C_{i})} = \int {\cal D}\Psi e^{-I[\Psi]} = e^{-I_{r}[\Psi]}+\cdots \,, 
\]
where $I_{r}[\Psi]$ denotes the on-shell renormalized action. By  writing the thermodynamic potential, ${\cal W}\equiv -\frac{1}{\beta}\ln Z$,  in terms of thermodynamic quantities as ${\cal W} = M-TS-\sum_{i}\mu_{i}C_{i}$,  one may say that the leading contribution to the thermodynamic potential ${\cal W}$ is related to the on-shell renormalized action value as
\begin{equation} \label{QSR}
 {\cal W} = M-TS-\sum_{i}\mu_{i}C_{i} =  \frac{1}{\beta} I_{r}[\Psi]\,.
\end{equation}
The above argument for the on-shell action value seems to be very plausible but  is not  self-evident,  especially in the theory of gravity. That is to say, the  quantum statistical relation should be taken with some caution, due to the fact that  the renormalization process has some intrinsic ambiguity unless we impose some renormalization condition. Furthermore, in the context of gravity, the characteristics of conserved charges are slightly different from those in  field theory  partially because charges in gravity are defined by the integration over the surface,  not the integration over the bulk. Even with these difficulties, the validity of the quantum statistical relation has been shown  in some specific cases~\cite{Gibbons:2004ai,Papadimitriou:2005ii}. 
In  Ref.~\cite{Papadimitriou:2005ii},  it was  shown that 
\begin{equation} \label{}
\frac{1}{\beta} I_{r}[\Psi] =   Q_{\infty}(\xi)-Q_{B}(\xi)\,,
\end{equation}
where $Q(\xi)$ denotes the charge for the Killing vector $\xi=\partial_t +\Omega\partial_\phi$.  See the Lemma 5.1 in  Ref.~\cite{Papadimitriou:2005ii}. 
For the stationary black holes in the asymptotic AdS space,  the above equation becomes
\begin{equation} \label{QSR}
\frac{1}{\beta}I_{r}[\Psi] = M -  T_{H}S-\Omega J - \mu Q\,.
\end{equation}
Furthermore, it was shown that the first law from the holographic computation can be completely matched with the Wald's bulk approach in the form of
\begin{equation} \label{First}
dM = T_{H}d{\cal S} + \Omega dJ +  \mu dQ\,.
\end{equation}
In this derivation of the quantum statistical relation, the (relaxed) Dirichlet boundary condition is chosen and, when the conformal anomaly function vanishes, the charge $Q_{\infty}(\xi)$ is shown to be composed only of the  metric and gauge fields parts. (See the Eq.~(4.31) in Ref.~\cite{Papadimitriou:2005ii}):
\[   
Q_{\infty}(\xi) = Q^{g}_{\infty}(\xi)  + Q^{A}_{\infty}(\xi)\,.
\]
Interestingly, this bulk result is also shown to be consistent with the boundary stress tensor method in~\cite{Papadimitriou:2005ii}, when the anomaly for the conformal symmetry is taken to vanish  under the (relaxed) Dirichlet boundary condition.

In contrast, it has been observed that the scalar hairy contribution could enter the charge $Q_{\infty}(\xi)$ in some AdS/CMT models~\cite{Gauntlett:2009dn,Gauntlett:2009bh,Sonner:2010yx,Dias:2013bwa} and  the corresponding hairy black holes~\cite{Hertog:2004ns,Henneaux:2006hk,Faulkner:2010fh}. Then, one may presume that all the hairy contributions could be incorporated into the mass or the angular momentum of hairy (planar) black holes and may retain the above quantum statistical relation Eq.~(\ref{QSR})  and the first law Eq.~(\ref{First}). However, there are various examples, which show the explicit hairy contribution to the thermodynamic relations. For instance, in Ref.~\cite{Gauntlett:2009dn,Gauntlett:2009bh}, the hairy contribution enters in the first law and this modified first law is used to explain the numerical results. These results invoke the questions as how we should improve the generic derivation of the quantum statistical relation and the first law. In the following section,  we show that all these results can be understood from the bulk side  by using the appropriate mass expression  and modifying the general derivation through the one-parameter path integration. Furthermore, we  confirm this result is consistent with the AdS/CFT correspondence from the boundary computation by using the improved boundary current. 

Before going ahead, it would be better to clarify the meaning of the scalar hairs and the boundary conditions of scalar fields. In our setup, especially in the context of the AdS/CMT correspondence, the scalar hairs correspond to free parameters of black holes just like the mass or angular momentum. Accordingly, we use the terminology of the hairy contributions from scalar fields  to stand for those from normalizable modes, not from non-normalizable modes which correspond to the fixed boundary values or the sources on the boundary. In planar black holes, both modes of the scalar fields in the asymptotic expansion of the radial coordinate could be  normalizable ones. Therefore, our question about the scalar hairy contribution corresponds to the role of these two modes in various thermodynamic relations under the Dirichlet boundary conditions\footnote{Note that the hairy contribution  corresponding to the multi-trace deformation could also be analyzed by the mixed boundary conditions~\cite{Papadimitriou:2007sj}.}.

\section{Revisit to the on-shell renormalized action}

Though there is a rigorous derivation of the quantum statistical relation of AdS black holes under the (relaxed) Diriclet boundary conditions~\cite{Papadimitriou:2005ii} (see Ref.~\cite{Papadimitriou:2007sj} for the approach in the case of the mixed boundary conditions), it does not cover  the class of hairy black hole models  in the AdS/CMT correspondence.  In this section we revisit thermodynamic relations of those black holes in a generic setup to clarify our claims and show that these cases could be interpreted nicely with our construction. In the following, we focus on the Dirichlet boudnary conditions, for definiteness.
 
\subsection{The case without the anomaly}

 The generic variation of the $D$ dimensional bulk action $I$ could be written as 
 \[   
 \delta I [\Psi]  =  \frac{1}{16\pi G}\int d^{D}x~ \delta ( \sqrt{-g}L ) =  \frac{1}{16\pi G}\int d^{D}x\Big[ \sqrt{-g}\CE_{\Psi}\delta \Psi + \partial_{\mu}\Theta^{\mu}(\delta \Psi)\Big]\,.
 \]
One may write the boundary  action in terms of  the surface integral at $\eta = \eta_{0}$ as
\[   
I_{GH}+ I_{ct} = \frac{1}{16\pi G}\int_{\eta=\eta_{0}} d^{D-1}x_{\mu}\, n^{\mu} \sqrt{-\gamma}(L_{GH}+L_{ct})\,. 
\]
where $n^{\mu}$ denotes the outward normal vector to the boundary surface at $\eta= \eta_{0}$. 
The on-shell Noether current  for the diffeomorphism parameter $\zeta$ of the bulk action $I$ could be introduced in the form of 
\begin{equation} \label{OnNoether}
J^{\mu} (\zeta) = \zeta^{\mu}\sqrt{-g} L  -\Theta^{\mu}(\Lie_{\zeta}\Psi)  = \partial_{\nu} K^{\mu\nu}\,,
\end{equation}
where $K^{\mu\nu}$ denotes the Noether potential and $\Theta^{\mu}$ is the surface term in the above action variation for the Lie derivative variation as $\delta \Psi = \Lie_{\zeta}\Psi$.  The  contribution of the Gibbons-Hawking and counter terms near infinity may  be incorporated by modifying the $\Theta$-term and the Noether potential as
\begin{align}   \label{ModTh}
\Theta^{\mu}(\delta \Psi)|_{\eta=\eta_{0}} &\longrightarrow ~~\tilde{\Theta}^{\mu}(\delta \Psi)|_{\eta=\eta_{0}} \equiv \Theta^{\mu}(\delta \Psi)|_{\eta=\eta_{0}} +  n^{\mu} \delta\Big[\sqrt{-\gamma}(L_{GH}+L_{ct})\Big]\,, \\   \label{ModK}
K^{\mu\nu}|_{\eta=\eta_{0}} &\longrightarrow ~~~~~~\tilde{K}^{\mu\nu}|_{\eta=\eta_{0}}  \equiv K^{\mu\nu}|_{\eta=\eta_{0}} +  2\zeta^{[\mu}_{B}\, n^{\nu]}\sqrt{-\gamma}(L_{GH}+L_{ct})\,,
\end{align}
where we have assumed that the generic variations do not change $n^{\mu}$, $\zeta^{\mu}$ and $\zeta^{i}_{B}$. 
One may note  that  the Noether potential $K^{\mu\nu}$ typically diverges in  the asymptotic AdS space and the counter terms  are introduced, by definition, to render it finite. More precisely, the counter terms should be considered as the essential ingredients to pose the well-defined variational problem~\cite{Papadimitriou:2005ii,Papadimitriou:2007sj}. In the explicit examples given in the following section, we adopt this property to determine the counter terms explicitly.
 
Now, let us recall that the Abbott-Deser-Tekin(ADT) potential \cite{Abbott:1981ff, Abbott:1982jh, Deser:2002rt, Deser:2002jk} for the bulk action $I$ is given by the combination of the Noether potential and the $\Theta$-term in the form of 
\begin{equation} \label{}
2\sqrt{-g} Q^{\mu\nu}_{ADT}(\xi) = \delta K^{\mu\nu}(\xi) -2\xi^{[\mu}\Theta^{\nu]}(\delta \Psi)\,,
\end{equation}
which gives us  to infinitesimal conserved charge for the Killing vector $\xi$ as 
\begin{equation} \label{ADTcharge}
\delta Q(\xi) = \frac{1}{8\pi G} \int d^{D-2}x_{\mu\nu}\, \sqrt{-g} Q^{\mu\nu}_{ADT}(\xi) 
\end{equation}
Remarkably, this charge expression can also be obtained by the covariant phase space method as the covariant Hamiltonian~\cite{Wald:1993nt,Iyer:1994ys}.  The existence of the improvement surface term of the form $\xi^{[\mu}\Theta^{\nu]}$ is now rather well understood and is important  to establish the first law of black holes through the Stokes' theorem.

As was emphasized in Ref.~\cite{Kim:2013zha,Kim:2013cor}, the ADT approach depends only on the bulk equations of motion. Henceforth, the conserved charge expression should be independent of the surface terms like the above Gibbons-Hawking, counter terms or additional boundary terms required for boundary conditions other than the Dirichlet ones. Indeed, by using Eq.~(\ref{ModTh}) and Eq.~ (\ref{ModK})  one can check explicitly that the modified Noether potential $\tilde{K}^{\mu\nu}$ and  $\tilde{\Theta}$ terms lead to the same ADT potential as 
\begin{equation} \label{ADTpotential}
2\sqrt{-g} \tilde{Q}^{\mu\nu}_{ADT}(\xi) = \delta \tilde{K}^{\mu\nu}(\xi) -2\xi^{[\mu}\tilde{\Theta}^{\nu]}(\delta \Psi) =  \delta K^{\mu\nu}(\xi) -2\xi^{[\mu}\Theta^{\nu]}(\delta \Psi) = 2\sqrt{-g} Q^{\mu\nu}_{ADT}(\xi)\,.
\end{equation}
We would like to emphasize that the boundary conditions are taken into account because the field variation in the above expressions means the variation along the one-parameter path in the solution space which respects the boundary conditions by construction.

Let us recapitulate the  results in Ref.~\cite{Papadimitriou:2005ii}, which are  relevant in our contexts~\cite{Hyun:2014sha}. 
In Einstein gravity, it was explicitly shown in Ref.~\cite{Papadimitriou:2005ii}  (see Eq.(3.46)) that the $\tilde{\Theta}$-term at $\eta=\eta_{0}$ becomes 
\begin{equation} \label{tildeTheta}
\tilde{\Theta}^{\mu}(\delta \Psi)\Big|_{\eta=\eta_{0}} = n^{\mu}\sqrt{-\gamma}\Big[ {\bf T}^{ij}_{B}\delta \gamma_{ij} + \Pi_{\psi}\delta \psi\Big]_{\eta=\eta_{0}}\,.
\end{equation}
By taking the conformal boundary condition such that the variation at the boundary should  be given by a Weyl transformation $\delta \rightarrow \delta_{\sigma}$ (See Eq.~(\ref{Weyl}) in Appendix), one can see that $  \tilde{\Theta}^{\mu}(\delta_{\sigma}\Psi) \sim  n^{\mu} {\cal A} \delta \sigma$, where ${\cal A}$ denotes the unintegrated anomaly~\cite{Henningson:1998gx,Papadimitriou:2005ii}. 
And thus  $\tilde{\Theta}^{\mu}$ vanishes, when the unintegrated anomaly ${\cal A}$ vanishes.

Notice that  the integrability condition to obtain   finite conserved charges  is  given in the form of~\cite{Wald:1999wa} 
\begin{equation} \label{IntCon}
0=\int_{\partial \Sigma} d^{D-2}x_{\mu\nu}\, \xi_{H}^{[\mu} \Big(\delta_{1}\tilde{\Theta}^{\nu]}(\delta_{2}\Psi) - \delta_{2}\tilde{\Theta}^{\nu]}(\delta_{1}\Psi) \Big)\,,
\end{equation}
which is satisifed automatically under the unintegrated anomaly vanishing condition ${\cal A}=0$.
Therefore, under the condition ${\cal A}=0$, the finite integrated   conserved charge could be determined just by $\tilde{K}$  as
\[   
Q(\xi_{B})  = \frac{1}{8\pi G}\int d^{D-2}dx_{\mu\nu}  \tilde{K}^{\mu\nu} (\xi_{B})  = \frac{1}{8\pi G}\int d^{D-2}dx_{\mu\nu} \Big[ K^{\mu\nu} (\xi_{B}) + 2\xi^{[\mu}_{B}\, n^{\nu]} \sqrt{-\gamma}(L_{GH}+L_{ct})\Big]\,.
\]
Moreover, it is straightforward to infer  from Eq.~(\ref{Bcur}) and Eq.~(\ref{BBequiv})  that this expression gives us the same charge expression from the boundary stress tensor ${\bf T}^{ij}_{B}$  as 
\begin{equation} \label{}
Q(\xi_{B}) =  \frac{1}{8\pi G}\int d^{D-1}x_{i}\, \sqrt{-\gamma} {\bf T}^{ij}_{B}\xi^{B}_{j}\,.
\end{equation}
In this case, hence, our construction is completely consistent with the results in Ref.~\cite{Papadimitriou:2005ii}. In Appendix, we show that the improved boundary current gives us the same results with the above bulk  expressions. 

Now, let us consider the relation between conserved charges and the renormalized on-shell action value. Firstly, let us recapitulate the derivation of the quantum statistical relation given in Ref.~\cite{Papadimitriou:2005ii}. 
Recalling the relation given in  Eq.~(\ref{OnNoether})   between the on-shell Noether potential and the Lagrangian value and using Eq.~(\ref{ModTh}) with $\delta=\pounds_{\xi}$, it is straightforward to  obtain   for the stationary Killing vector $\xi_{H} $ 
\begin{equation} \label{OnactionRel}
\int d^{D-1}x{\sqrt{-g}L} + \int d^{D-2} x_{i}\, n^{i}\sqrt{-\gamma}(L_{GH}+L_{ct}) = \bigg(\int_{\infty}- \int_{\cal B} \bigg)d^{D-2}x_{\mu\nu}\, \tilde{K}^{\mu\nu} (\xi_{H})\,.
\end{equation}
The left hand side of this equality is nothing but the on-shell renormalized Lagrangian integrated over the relevant domain except the Euclidean time integration and the  right hand side corresponds to the conserved charges at infinity and on the horizon,  when the unintegrated trace anomaly vanishes. If the gauge is chosen appropriately, the conserved charge at infinity and on the horizon may be identified, respectively, as
\begin{align}   \label{}
Q_{\infty}(\xi_{H}) &= \frac{1}{16\pi G}\int_{\infty}d^{D-2}x_{\mu\nu}\, \tilde{K}^{\mu\nu} (\xi_{H}) = M- \Omega J - \mu Q\,, \nn  \\
  Q_{\cal B}(\xi_{H})  &= \frac{1}{16\pi G} \int_{\cal B}d^{D-2}x_{\mu\nu}\, \tilde{K}^{\mu\nu} (\xi_{H}) = T_{H}{\cal S}\,,  \nn
\end{align}
where the conserved charge on the horizon is identified with the black hole entropy {\it a la} Wald~\cite{Wald:1993nt,Iyer:1994ys}.  For the stationary  system in the adapted coordinates in such a way that $\xi^{t}_{H} =1$, the Euclidean time integration can be performed simply by multiplying its interval and thus  the above  equality for the Killing vector $\xi_{H}$ leads to 
\begin{equation} \label{}
\frac{1}{\beta} I_{r} = M - \Omega J -\mu Q - T_{H}{\cal S} \,,
\end{equation}
which is nothing but  the statement of the Lemma 5.1 {\it i)} in Ref.~\cite{Papadimitriou:2005ii}. In the following section, we would like to explore the case where the unintegrated anomaly does not vanish.

\subsection{The case with the anomaly}

In this section, we would like to consider cases in which  the unintegrated anomaly ${\cal A}$ does not vanish under the Drichlet boundary conditions.  In fact, there is an approach allowing  the non-vanishing anomaly function  which  incorporates it into the conserved current~\cite{Hyun:2014sha}.   Furthermore, the equivalence of the bulk and boundary conserved charges is established through Eq.(54)  and (55) in Ref.~\cite{Hyun:2014sha}. The essential point in this construction is that the ambiguity in the counter terms does not affect the holographic charges, since the improved boundary currents are matched with the bulk ADT potentials.
Our suggestion in  the case of non-vanishing anomaly  is to use the improved boundary current for the boundary conserved charge and to use the conventional bulk expression or the ADT potential for the bulk conserved charge. The equivalence of these two computations is shown in Ref.~\cite{Hyun:2014sha} and thus it is completely consistent with the AdS/CFT correspondence.  Concretely, one can define the infinitesimal conserved charge  as 
\begin{equation} \label{}
\delta Q(\xi_{B}) = \frac{1}{8\pi G} \int_{\eta=\eta_{0}} d^{D-2}x_{\mu\nu}~ \sqrt{-g} Q^{\mu\nu}_{ADT}(\xi) =\frac{1}{8\pi G} \int d^{D-2}x_{i}\, \sqrt{-\gamma}{\cal J}^{i}_{B}(\xi_{B})\,,  
 \end{equation}
and then  integrate this along the one-parameter path in the solution space. 
 In the next section, we apply this suggestion to the specific models and confirm that the thermodynamic relation may be explained consistently and still the renowned first law of black holes is retained without any modification. Furthermore, the quantum statistical relation could be retained in its form by taking the counter terms appropriately.

Now, let us extend the derivation of the quantum statistical relation when the unintegrated trace anomaly does not vanish.
The on-shell  variation of the relation~(\ref{OnactionRel}) leads to
\begin{align}   \label{}
\delta \Big(\frac{1}{\beta}I_{r}\Big) &= \frac{1}{16\pi G}\bigg(\int_{\infty}- \int_{\cal B} \bigg)d^{D-2}x_{\mu\nu}~  \delta \tilde{K}^{\mu\nu} (\xi_{H})  \nn \\
& = \frac{1}{8\pi G}\bigg(\int_{\infty}- \int_{\cal B} \bigg)d^{D-2}x_{\mu\nu}~   \sqrt{-g}Q^{\mu\nu}_{ADT}   +  \frac{1}{8\pi G}\bigg(\int_{\infty}- \int_{\cal B} \bigg)d^{D-2}x_{\mu\nu}~ \xi^{[\mu}_{H}\tilde{\Theta}^{\nu]}(\delta \Psi)\,, \nn 
\end{align}
where we have used Eq.~(\ref{ADTpotential}) to rewrite $\delta\tilde{K}^{\mu\nu}$ in terms of $Q_{ADT}^{\mu\nu}$ and $\tilde{\Theta}^{\nu}$.
By using the infinitesimal form of conserved charges given by Eq.~(\ref{ADTcharge}) and using the property $\xi_{H}\rightarrow 0$ on  the bifurcation horizon, we obtain the following result:
\begin{equation} \label{}
\delta \Big(\frac{1}{\beta}I_{r}\Big) = \delta Q_{\infty}(\xi_{H}) -  \delta Q_{\cal B}(\xi_{H}) +  \frac{1}{8\pi G}\int_{\infty}d^{D-2}x_{\mu\nu}~ \xi^{[\mu}_{H}\tilde{\Theta}^{\nu]}(\delta  \Psi)\,.
\end{equation}

As a result, one can see that  additional terms may exist at infinity such as  the last term in the above equality.  Since we should take the specific one-parameter path in the solution space for this on-shell variation and  should define conserved charges consistently, we need to impose the integrability condition on the allowed configurations. 
Under this assumption, one can integrate the last term and obtain the following  finite form of the quantum statistical relation
\begin{equation} \label{modQSR}
\frac{1}{\beta}I_{r} = M - \Omega J  - \mu Q -T_{H}{\cal S} +  \frac{1}{8\pi G}\int_{\infty}d^{D-2}x_{\mu\nu}~ \xi^{[\mu}_{H}\tilde{\cal B}^{\nu]}(\delta \Psi)\,,
\end{equation}
where we have assumed that $\xi_{H}$ does not change along the one-parameter path and $\tilde{B}^{\mu}$ denotes the integrated expression of $\tilde{\Theta}^{\mu}$ along the one-parameter path in the solution space: $\tilde{\cal B}^{\mu} \equiv \int ds \tilde{\Theta}^{\mu}(\delta_{s}\Psi)$.  
Since we are considering the Dirichlet boundary conditions, the on-shell solutions or the one-parameter path in the solution space should respect these boundary conditions~\cite{Anabalon:2015xvl,Papadimitriou:2005ii,Papadimitriou:2007sj}, which means that
\begin{align}
	\lim_{\eta_{0}\rightarrow \infty} \delta_{s}\Psi \Big|_{Dirichlet} =0\,,
\end{align}
where $\eta_{0}$ denotes the position of the boundary surface which is sent to infinity at the end.
By recalling the generic expression of $\tilde{\Theta}^{\mu}(\delta_{s}\Psi)$ given in Eq.~(\ref{tildeTheta}), which shows that it depends linearly on $\delta_{s}\Psi$, its integrated form ${\cal \tilde{B}}^{\mu}$ should vanish as far as the integrability condition holds.   As a result, under the Dirichlet boundary conditions, we obtain
\begin{equation} \label{}
\frac{1}{\beta}I_{r} = M - \Omega J  - \mu Q -T_{H}{\cal S}\,,
\end{equation}
which is our final result on the quantum statistical relation.

 In the next section, we revisit the models in Ref.~\cite{Gauntlett:2009dn,Gauntlett:2009bh} to illustrate  the power of our formulation.   In this class of models, one can clearly see that  the unintegrated anomaly function does not vanish in general.  Therefore, the generic proofs for the quantum statistical relation and the first law in Ref.~\cite{Papadimitriou:2005ii}, in which the vanishing of anomaly function is assumed, would be insufficient to cover  this class of models. Though there is an improved approach covering these hairy models through the mixed boundary conditions~\cite{Papadimitriou:2007sj}, it would be interesting to analyze these models by using the Dirichlet boundary conditions just as in Ref.~\cite{Gauntlett:2009dn,Gauntlett:2009bh,Sonner:2010yx}, where the first law of black holes is modified appropriately,  while the quantum statistical relation is retained in its form.  In these references, the modification of the first law is given by hand, rather than is based on a definite formalism. And, the relevant thermodynamic quantities are evaluated only from the boundary side and are not matched with expressions from the bulk computation.

 One may guess that our bulk formulation  and its implementation  to thermodynamic relations presented in the above  may be in conflict to Ref.~\cite{Gauntlett:2009dn,Gauntlett:2009bh}.  
However, as will be shown explicitly in the next section,  this does not mean the conflict  with the numerical results. The essential modification resides in the on-shell renormalized action value, which depends on the choice of counter terms. It just gives  the consistent reinterpretation of those results based on a definite formalism without any {\it ad hoc} modification.  Moreover, we would like to emphasize  that our formulation is completely consistent with the modification of the boundary stress tensor taken in the model for the boundary vortices in Ref.~\cite{Dias:2013bwa} (see also Ref.~\cite{Domenech:2010nf, Montull:2012fy, Salvio:2012at, Salvio:2013jia}). In that work,  in order to overcome the non-conservation of the covariant derivative of the boundary stress tensor,  it was suggested  to use the boundary stress tensor modified by adding the appropriate expression of the  condensate and then the consistency  was checked with the first law in the standard form as given in Eq.~(\ref{First}). Our  formulation gives the same conclusion with this modification, as can be verified straightforwardly.

\section{Holographic models}
In this section, we focus on a specific model and present detailed expressions to  illustrate  the general arguments given in the previous section.   We will see that all the seemingly conflicting results in the literatures are resolved naturally in our formulation. We consider the model  for  the holographic superconductor  embedded in M-theory~\cite{Gauntlett:2009dn}. It contains the non-minimally coupled  complex scalar and the $U(1)$ gauge fields whose action is given by
\begin{equation} \label{}
I[g,{\cal A},\varphi] =\frac{1}{16\pi G}\int d^{4}x\sqrt{-g}\,\bigg[  R - \frac{1}{4}{\cal F}_{\mu\nu}{\cal F}^{\mu\nu} + \frac{1}{(1-\frac{1}{2}|\varphi|^2)^2} \Big( -|{\cal D_\mu}\varphi|^2 + 24\big( 1-\frac{2}{3}|\varphi|^2 \big) \Big) \bigg]\,,
\end{equation}
where ${\cal D_\mu}\varphi\equiv \partial_\mu \varphi -2iA_\mu \varphi$.
The metric,  the complex scalar and the $U(1)$ gauge fields ansatz are taken as
\begin{align}   \label{ansatz}
ds^{2}&= -e^{2A(r)}f (r)  dt^{2} + \frac{dr^{2}}{f(r)} +r^{2}d{\bf x}^{2}\,, \\
\varphi &= \sigma(r)\in\mathbb{R} \,, \qquad {\cal A} = A_{t}(r)dt\,. 
\end{align}
The asymptotic expansions of those fields are given by
\begin{align}   \label{}
A(r) &= -\frac{\beta_a}{2} -\frac{\sigma_1{}^2}{4r^2} -\frac{2\sigma_1\sigma_2}{3r^3} \cdots\,,\\
f(r) &= 4 r^{2} + 2\sigma_1{}^2 -\frac{m-4\sigma_1\sigma_2}{r} +\cdots\,, \nn\\
\sigma(r) &= \frac{\sigma_1}{r} +\frac{\sigma_2}{r^2} + \cdots\,, \qquad A_t(r) = e^{-\beta_a/2}\Big( \mu-\frac{q}{r} \Big)+\cdots\,. \nn
\end{align}
%
%
%
%
By using the symmetry of solutions, one can take $\beta_a=0$ without loss of generality. (See the Eqs.(5.6) and (5.7) in Ref.~\cite{Gauntlett:2009bh}.) In the following, we take $\beta_a=0$ for the standard normalization of the metric.
In order to preserve the asymptotic AdS structure, we need to take the relation between two asymptotic scalar values as
\begin{align} \label{ScalarModRel}
	\sigma_{2}= \nu\sigma^{2}_{1}\,,
\end{align}
where $\nu$ denotes a dimensionless constant~\cite{Hertog:2004ns}. 
Though this relation may be regarded as the relation between the boundary value and its momentum value in the context of the mixed boundary approach to the hairy contribution, we would like to take this just as the relation between two free parameters since both of $\sigma_{1}/r$ and $\sigma_{2}/r^{2}$ correspond the normalizable modes. 

As alluded in the previous section, the appropriate choice of counter terms is essential to obtain the well-defined variational problem under the Dirichlet boundary condition. Explicitly, we impose the following condition for the variation of the on-shell action  in order to obtain a well-posed  variational problem  under the Dirichlet boundary conditions 
\begin{equation} \label{condition}
\delta I_{r}[\Psi]\Big|_{Dirichlet} = 0\,, \qquad I_{r} = I + I_{GH}+I_{ct}\,,
\end{equation}
where $I_{GH}$ denotes the Gibbons-Hawking term and $I_{ct}$ does the counter term. Note that the conventional choice of counter term for the scalar field, $I_{ct}[\sigma] \sim \sigma^{2}$, in~\cite{Gauntlett:2009dn,Gauntlett:2009bh} is not compatible with the above condition. Our choice of the counter term for the scalar field $\sigma$ which satisfies the above condition is\footnote{Note that  either the  conventional choice, $I_{ct}[\sigma] \sim \sigma^{2}$ or our choice of counter terms gives us the finite renormalized action up to the finite difference. Therefore, the finiteness of the renormalized action is not sufficient to choose the counter terms completely in this case (see also~\cite{Anabalon:2015xvl}).} 
\begin{equation} \label{Counter}
I_{ct}[\sigma]= \frac{1}{16\pi G} \int d^3 x \sqrt{-g} \Big[-2\sigma^2 -\frac{4\nu}{3} \sigma^3 \Big] \,.
\end{equation}
%

\subsection{Revisit to thermodynamic relations: Bulk side}
Now, let us compute the mass of these black holes by using the quasi-local formalism~\cite{Kim:2013zha,Kim:2013cor,Hyun:2014kfa,Hyun:2014nma,Hyun:2016dvt} which is completely equivalent to the covariant phase space method~\cite{Wald:1993nt,Iyer:1994ys}.   By using the formula for conserved charges given in Eq.~(\ref{ADTcharge}),
the infinitesimal expression of the black hole mass, corresponding to  the Killing vector $\xi=\partial_t$,  can be obtained as
\begin{equation} \label{}
\delta M|_{r\rightarrow\infty} = \frac{V_2}{8\pi G}\left[ -e^A r \delta f - \frac{e^A r^2 f\sigma'}{\big(1-\frac{1}{2}\sigma^2\big)^2} \delta\sigma \right]_{r\rightarrow\infty} = \frac{V_2}{8\pi G}\Big[\delta m +4\sigma_2\delta\sigma_1\Big]\,,
\end{equation}
where $V_2$ denotes the $2$-dimensional volume element.
By integrating over the parameters $m$ and $\sigma_1$ along with the relation (\ref{ScalarModRel}),   the finite bulk expression of the  mass is  given by
\begin{align}\label{mass}
	M = \frac{V_2}{8\pi G}\Big[ m + \frac{4}{3}\nu\sigma_1^3 \Big]
= \frac{V_2}{8\pi G}\Big[m+ \frac{4}{3}\sigma_{1}\sigma_{2} \Big]\,.
\end{align}
The temperature, the black hole entropy and the $U(1)$ charge are given by the surface gravity $\kappa$, by the area law or by the Wald formula and by a conventional form, respectively, as
\begin{equation} \label{}
T_H\equiv\frac{\kappa}{2\pi}=\frac{1}{4\pi}\,  e^{A(r_{H})}f'(r_{H})\,,\qquad {\cal S}\equiv \frac{V_2{\hat s}}{8\pi G}  = \frac{r^2_{H}V_2}{4G} \,, \qquad Q = \frac{qV_2}{8\pi G}\,.
\end{equation}
By using the above mass expression, one can see that the first law  takes the form
\begin{equation} \label{FirstLaw}
dM = T_{H}d{\cal S} +  \mu dQ\,,
\end{equation}
which is irrespective of the explicit form of counter terms,
while the quantum statistical relation is given by
\begin{equation} \label{}
\frac{1}{\beta}I_{r} = M  -T_{H}{\cal S} - \mu Q \,,
\end{equation}
 for the above chosen counter terms in Eq.~(\ref{Counter}).
By using the mass expression given in Eq.~(\ref{mass}), one can see that 
\begin{align}
\frac{1}{\beta}I_{r} = \frac{V_2}{8\pi G}\Big[ m+\frac{4}{3}\sigma_1\sigma_2  -T_{H}{\hat s}  - \mu q \Big]\,,	
\end{align}
which is different from Ref.~\cite{Gauntlett:2009dn,Gauntlett:2009bh} just by the finite piece. That is to say, 
the difference in counter terms gives us just the finite difference between our expression and Ref.~\cite{Gauntlett:2009dn,Gauntlett:2009bh}.


One can generalize the relation between two modes of the scalar field as in Ref.~\cite{Hertog:2004dr,Hertog:2004bb,Hertog:2004ns}. In order to assure that the integrability condition in Eq.~(\ref{IntCon}) hold, let us take 
\[   
\sigma_{2} = H(\sigma_{1})\,,
\]
and assume that the log-mode does not appear in this expansion. 
%
%
We should be careful in choosing the counter terms in order to respect the Dirichlet boundary conditions. A simple computation reveals  that appropriate counter term for scalar field is given by $L_{ct}({\sigma})=-2\sigma^2 - \frac{4}{3\sigma_1^2} H(\sigma_{1})\, \sigma^{3}$.
%
For this choice of counter terms,  one can easily  verify  our claim that 
\begin{equation} \label{}
\tilde{\Theta}^{\eta} (\delta \sigma)_{\eta_{0}\rightarrow\infty} = 0\,.
\end{equation}
 Then, one obtains immediately the following expression
\begin{equation} \label{}
  \frac{1}{8\pi G}\int_{\infty}d^{2}x_{\mu\nu}~ \xi^{[\mu}_{H}\tilde{\cal B}^{\nu]}(\delta \Psi) = 0\,.
\end{equation}
In Ref.~\cite{Hertog:2004ns,Faulkner:2010fh}, it was shown that the asymptotic AdS algebra structure is preserved only when the  two modes are related by the dimensionless parameter $\nu$ as $H(\sigma_{1}) = \nu \sigma_{1}{}^{2}$. Interestingly, the quantum statistical relation and the first law could hold in the usual forms under the generalized relation between $\sigma_{1}$ and $\sigma_{2}$.

Now, some comments are in order.
 In the above  example, one may think that  the first law of black holes  given in Eq.~(\ref{FirstLaw})  is the  simple rewriting of  the corresponding expressions in Ref.~\cite{Gauntlett:2009dn,Gauntlett:2009bh} just by redefining the mass expression. To avoid such impression, we would like to emphasize  our main points as follows.  The mass expression given in the above is not an {\it ad hoc} redefinition  but  is the  consequence of quasi-local ADT formalism or the covariant phase space approach for conserved charges, which are well-established in the bulk gravity side. Furthermore, in the following section we will show that the same mass expression can be obtained  from the improved boundary current in the framework of the holographic renormalization.  At the level of a generic model,  our construction and the resultant expression reveal how to generalize the derivation of the quantum statistical relation and the first law  given in~\cite{Papadimitriou:2005ii}, even with the non-vanishing unintegrated trace anomaly in the boundary by using the Dirichlet boundary conditions. Our results would be consistent with the analysis by using the mixed boundary conditions~\cite{Papadimitriou:2007sj}. Moreover, for planar AdS black holes, we will show that the Smarr-like relation could be obtained consistently with our claimed quantum statistical relation and the first law.

\subsection{Revisit to thermodynamic relations: Boundary side}
We would like to summarize the interesting properties of our improved boundary current given in Appendix.  First, one can see that the first and the second  term in Eq.~(\ref{Bcur}) are conserved separately and so this current could be used  to define conserved charges for the boundary Killing vector $\xi_{B}$. One may recall that the first term or its integrated version is adopted in the conventional holographic computation, while the second term for the Weyl transformation corresponds to the anomaly function on the boundary that should be taken to vanish to preserve  the conformal symmetry.   Even for the non-vanishing conformal anomaly function under the Dirichlet boundary conditions, the second term in Eq.~(\ref{Bcur}) could vanish. Then,  it should be dropped from the expression. However, we need to choose the counter terms given in Eq.~(\ref{Counter}) in order to achieve this goal.  The power of the improved current construction resides in the fact that its equivalence to the bulk ADT potentials, which allows us to use these currents even for the conventional counter term $L_{ct}=-2\sigma^2$. In the following, we present some explicit computations to verify these claims.

By using the improved boundary current given in Appendix, one can compute the infinitesimal mass expression as
\begin{align}   \label{}
\delta M &= \delta Q(\xi^{T}_B) = \frac{1}{8\pi G}\int d^{2}x \,n_{i}  \sqrt{-g}{\cal J}^{i}_{B}  \nn \\
& = \frac{1}{8\pi G}\int d^{2}x \,n_i \Big[-\delta \Big(\sqrt{-\gamma}{\bf T}^{~i}_{\!B\, j}\xi^{j}_{\!B}\Big)  + \frac{1}{2}\sqrt{-\gamma}\xi^{i}_{B} \Big(T^{kl}_{B}\delta \gamma_{kl} + \Pi_{\psi}\delta \psi\Big)\Big]\,.
\end{align}
Here, we need adopt the Fefferman-Graham expansion for the complete consistency with the bulk results:
\begin{align}
		ds^2 &= d\eta^2+\gamma_{ij}dx^i dx^j \,,\\
	\gamma_{tt} &= 4 e^{4\eta} \Big[ -1+\frac{\sigma_1{}^2}{4}e^{-4\eta} +\frac{m+4\sigma_1 \sigma_2}{6}e^{-6\eta} +\cdots \Big]\,,\\
	\gamma_{xx} = \gamma_{yy} &= e^{4\eta} \Big[ 1-\frac{\sigma_1{}^2}{4}e^{-4\eta} +\frac{m-4\sigma_1 \sigma_2}{12}e^{-6\eta} +\cdots \Big]\,.
\end{align}
The explicit computation with the proper counter terms for the model in Eq.~(\ref{Counter})  shows us that the each term for the boundary current is
%
\begin{align}
\frac{1}{8\pi G}\int d^{2}x \,n_i \Big[-\delta \Big(\sqrt{-\gamma}{\bf T}^{~i}_{\!B\, j}\xi^{j}_{\!B}\Big)  \Big] &=\frac{V_2}{8\pi G} \, \Big[\delta m  + 4\sigma_2\delta\sigma_1 \Big] \,,\\
\frac{1}{8\pi G}\int d^{2}x \,n_i \Big[ \frac{1}{2}\sqrt{-\gamma}\xi^{i}_{B} \Big(T^{kl}_{B}\delta \gamma_{kl} + \Pi_{\psi}\delta \psi\Big)\Big]&=0  \,.
\end{align}
By using the relation (\ref{ScalarModRel}), one can integrate  above expression to obtain
\[   
M = \frac{V_2}{8\pi G} \Big[m + \frac{4}{3}\sigma_{1}\sigma_{2}\Big]\,,
\]
which is   the same   as the bulk expression. 

 Now, let us choose  counter terms as  $L_{ct}=-2\sigma^2$, which gives us
\begin{align}   \label{}
\frac{1}{8\pi G}\int d^{2}x \,n_i \Big[-\delta \Big(\sqrt{-\gamma}{\bf T}^{~i}_{\!B\, j}\xi^{j}_{\!B}\Big)  \Big] &=\frac{V_2}{8\pi G} \, \delta m   \,,\\
\frac{1}{8\pi G}\int d^{2}x \,n_i \Big[ \frac{1}{2}\sqrt{-\gamma}\xi^{i}_{B} \Big(T^{kl}_{B}\delta \gamma_{kl} + \Pi_{\psi}\delta \psi\Big)\Big]&=\frac{V_2}{8\pi G} \,  4\sigma_2\delta\sigma_1  \,.
\end{align}
Indeed, one can see that the mass expression becomes identical with the one in the above.

\subsection{The Smarr-like relation}

In the case of the Kerr-AdS black holes, it was known~\cite{Gibbons:2004ai} that the Smarr(-Gibbs-Duhem) relation does not hold in general, while the first law of black holes holds in its universal form.   However, it has been shown~\cite{Ahn:2015shg} that the Smarr-like relation can be obtained in the asymptotic AdS planar black holes model-independently. 

Our starting point is to use the ansatz (\ref{ansatz}) to obtain the reduced action:
\begin{equation} \label{reduced action}
I_{red}[\tilde\Psi] = \frac{1}{16\pi G} \int dt d{\bf x}\, \int dr \,L_{red}(r, \tilde\Psi )\,,
\end{equation}
where  $\tilde\Psi$  denotes collectively all the functions of the radial coordinate $r=e^{\eta}$  appearing in the ansatz of all the fields.
The corresponding reduced Lagrangian for our model is given by 
\begin{align}
L_{red}=-e^A(2rf)' + \frac{1}{2}r^2 e^{-A}A_t{}'^{\,2} + \frac{r^2\Big[- e^{A}f\sigma'{}^2 + 4e^{-A}f^{-1}\sigma^2A_t{}^2 + 24 e^{A}(1-\frac{2}{3} \sigma^2) \Big]}{ \Big(1-\frac{1}{2}\sigma^2 \Big)^2}\,,
\end{align}
where  ${}'$ denotes the derivative with respect to the coordinate $r$.
As has been shown in \cite{Hyun:2015tia,Ahn:2015uza,Ahn:2015shg}, the reduced action has  the off-shell scaling symmetry 
under the rescaling of the radial coordinate $r$, if we assign the reduced fields $\tilde\Psi$ to transform appropriately, with the definite weight, under the transformation. 
One can apply the standard Noether method to obtain the  associated charge as~\cite{Hyun:2015tia,Ahn:2015uza,Ahn:2015shg}
\begin{align}
	c(r)= \frac{V_2}{16\pi G} \bigg[ 2e^A r(rf'-2f) -2e^{-A}r^2 A_t A_t{}' +\frac{2e^A r^3 f\sigma'{}^2+8e^{-A}r^3 f^{-1}\sigma^2A_t{}^2}{\big(1-\frac{1}{2}\sigma^2 \big)^2} \bigg]\,.
\end{align}
Note that the charge $c$ is invariant along radial direction. It seems natural to expect that its expression at the asymptotic infinity would be related to those of  physical quantities defined at the asymptotic infinity. Indeed it is given by
\begin{align}
c= c(r\rightarrow\infty)= 	\frac{V_2}{8\pi G}   \Big[3m+4\sigma_1\sigma_2-\mu q\Big] = 3M - 2\mu Q\,. 
\end{align}
On the other hand, the expression of the charge can be computed on the horizon as
\begin{align}
c = c(r=r_{H}) = \frac{r^{2}_{H}V_2}{8\pi G}\,  e^{A(r_{H})}f'(r_{H})= 2 T_{H} {\cal S} \,. 	
\end{align}
Thus we obtain the Smarr-like relation:   
\begin{equation} \label{}
c =  2 T_{H} {\cal S}    =3 M - 2 \mu Q\,.
\end{equation}
Note that there is no hairy contribution in this relation and this is completely consistent with the first law and the quantum statistical relation. Since we are considering the homogeneous system in the dual field theory side, the pressure of the dual system is simply given by the thermodynamic potential in the grand canonical ensemble. 
Combining the quantum statistical relation given in (\ref{mass}) with the above Smarr-like relation, one can see that the pressure of the dual system is given by
\begin{equation} \label{}
T_H I_r = -{\cal P} = M  -T_{H}{\cal S} - \mu Q = -\frac{1}{2}M  \,.
\end{equation}
Then, this pressures  could also be rewritten  in the form of 
\[   
{\cal P} = \frac{V_2}{8\pi G}\Big[ \frac{1}{2}m +\frac{2}{3}\sigma_1\sigma_2 \Big]= -\frac{V_2}{8\pi G}\Big[ m+\frac{4}{3}\sigma_1\sigma_2 - \mu q - T_{H}\hat{s}\Big]\,,
\]
which is consistent with the first law and the quantum statistical relation.

\section{Conclusion}

In asymptotic AdS space, black holes could have non-trivial hairs which have  interesting interpretation in the context of the AdS/CMT correspondence. Specifically, the existence of non-trivial scalar hairs is deeply related to  the holographic realization of the superconductors or superfluids. More concretely, double trace deformations in the dual field theory are related to turning-on of two kinds of normalizable modes in the bulk. In the context of the AdS/CMT models, the simultaneous turning-on of those two modes implies the condensate in the dual field theory. In this situation the unintegrated trace anomaly cannot vanish. Hence, the thermodynamic relations may be modified from the standard form. Indeed, various AdS/CMT models have been constructed to reveal such modification but the interpretation of the modification does not seem to have consensus.

 In this paper, we have revisited the thermodynamic relations in the hairy AdS planar black holes. Because of non-trivial scalar hairs,  one may guess that the standard form of the quantum statistical relation and the first law can not hold simultaneously. Based on the consistent bulk/boundary formalism,  we have shown that one can retain the first law as its universal form given in Eq.~(\ref{First}) and  the quantum statistical relation is not modified  as given in Eq.~(\ref{modQSR}). In summary, the first law, the quantum statistical relation and the Smarr-Gibbs-Duhem relation do not contain the explicit hairy contribution but has the implicit contribution hidden in the conserved charges.  It was noticed by Wald~\cite{Wald:1993nt,Iyer:1994ys,Wald:1999wa} that  the existence of the additional term beyond the Komar integrand in the expression of conserved charges is crucial to resolve the factor two difference between the mass and angular momentum expression of black holes and to establish the first law of black holes. As shown in this paper,  the additional term to the boundary current  gives us the results independent of counter terms, which is natural from its matching with the bulk side expressions. We have shown that the seemingly clashing expressions in the boundary computation could be improved by using the improved boundary current in the framework of the holographic renormalization.

By revisiting the explicit examples  and taking the Dirchlet boundary conditions, we have checked our formulation and shown the consistency of our interpretation. We have also commented the integrability issues to define conserved charges. Concretely, it has been well known that the integrability issue exists in the covariant phase space approach to define conserved charges in the bulk perspective. It is natural to anticipate in the context of the AdS/CFT correspondence that the same issue would appear in the boundary computation if holographic charges can be completely identified with those from covariant phase space approach. Interestingly, this issue could be ignored when the unintegrated trace anomaly vanishes. However, as is evident in the various AdS/CMT models, there are consistent models with the non-vanishing unintegrated anomaly. Hence, the integrability issue should appear in the boundary computation if we try to identify  holographic charges  with bulk charges in the covariant phase space approach. Indeed, we have verified that  this is the case and the modified boundary current contains such information. One may recall that  Kerr-AdS black holes are another examples~\cite{Hyun:2014nma} showing the similar phenomenon for the complete  identification between the bulk and the boundary charges. 

It would be interesting to investigate  other examples to check our claims and to see the meaning of the non-vanishing unintegrated anomaly in other contexts. Specifically, it would be interesting to consider rotating hairy black holes to see their thermodynamic relations. It would be very interesting to see whether the thermodynamic (in)stability criterion would be changed or not  by our thermodynamic relations.  Since some part of conformal symmetry would be broken in the non-vanishing anomaly, it would be interesting to see how much the AdS/CFT or AdS/CMT correspondence tells us the matching between the bulk and boundary quantities.

\vskip 1cm
\centerline{\large \bf Acknowledgments}
\vskip0.5cm
{We would like to thank  Byoungjoon Ahn, Jaehoon Jeong and Kyung Kiu Kim for some discussion.
SH was supported by the National Research Foundation of Korea(NRF) grant 
with the grant number NRF-2013R1A1A2011548. SY was supported by the National Research Foundation of Korea(NRF) grant with the grant number NRF-2015R1D1A1A09057057.}
%

{\center \section*{Appendix: Improved boundary currents and conserved charges}}

\renewcommand{\theequation}{A.\arabic{equation}}
  \setcounter{equation}{0}
 
In this appendix, we summarize on the improved boundary current and its relation to the bulk ADT potential.  Though we have shown in the main text through the bulk computation that the hairy contribution to black holes may be incorporated into  conserved charges, while  the quantum statistical relation, the first law of black holes and the Smarr-like relations hold in the usual forms, it would be more satisfactory that the same conclusion could be achieved consistently in the framework of the holographic renormalization. Since the relevant formulation for this improved boundary currentis already given in~\cite{Hyun:2014sha,Hyun:2014nma}, we present the summary of relevant stuffs in the following for the completeness.  As is alluded in the main text, the additional term in the improved boundary current could vanish  with the appropriate counter terms {\it under the Dirichlet boundary conditions}, while the existence of the additional term in the boundary currents allow us to use those even with the usually adopted form of counter terms. 
%
%

Before going into some details, it would be better to mention our motivation to construct the improved boundary current. At the superficial level, the bulk expression for conserved charges in the quasi-local ADT formulation or the covariant phase space approach could be defined completely independent of the boundary terms, while the conventional holographic expressions for conserved charges depend, at least weakly, on the chosen counter terms. In order to avoid such mismatch and to  warrant the complete match between bulk and boundary expressions, we have proposed in Ref.~\cite{Hyun:2014sha} to improve the  current  expression for  boundary conserved charges, while keeping the conventional bulk expression  for  bulk conserved charges in covariant phase space approach or equivalently in the quasi-local ADT formalism. 
 
From the bulk perspective, the vanishing unintegrated anomaly is not an essential requirement in order to define conserved charges for Killing vectors. Rather,   the integrability  of infinitesimal conserved charges along the on-shell  parameter space is more relevant  issues in defining well-defined conserved charges (See, for instance Ref.~\cite{Wald:1999wa}).  In other words, the vanishing anomaly condition is consistent but not mandatory in the bulk side, as can be inferred from the fact that  the vanishing unintegrated anomaly implies the integrability of infinitesimal bulk charges but not vice versa.   Therefore, one may consider  relaxing the condition that  the  unintegrated trace anomaly  vanishes at the boundary  even under the Dirichlet boundary conditions (see~\cite{Papadimitriou:2007sj} for a different approach). Furthermore, some models in the AdS/CMT correspondence realize such cases.  In fact, it was already noticed in Ref.~\cite{Papadimitriou:2005ii} that conserved charges could be defined for Killing vectors (but not conformal Killing ones), even when the unintegrated anomaly does not vanish.  In this case, one may introduce the improved boundary current in the context of the AdS/CFT correspondence, which is suitable for  relaxing the condition that the unintegrated trace anomaly  vanishes at the boundary.  Then, one can obtain the completely consistent bulk/boundary conserved charges.  
 
 In the following, we adopt the definition of conserved charges through the covariant phase space approach or equivalently the quasi-local ADT method in the bulk. At the boundary,  we use the charge expression obtained by using the improved boundary current, not  just the conventional holographic current expression which is constructed through the boundary stress tensor by contracting it with the boundary Killing vector.  It turns out that various thermodynamic relations  can be interpreted consistently for various  models of the non-vanishing unintegrated trace anomaly  under the Dirichlet boundary conditions. 

To introduce the modified boundary current, let us consider the following Fefferman-Graham expansion for the asymptotically AdS spacetime
\[   
ds^{2} = d\eta^{2} + \gamma_{ij} dx^{i}dx^{j}\,,
\]
where the time-like boundary of the AdS space is taken at $\eta = \eta_{0} \rightarrow\infty$.  In asymptotically AdS space, the renormalized on-shell action can be written  by  introducing the counter term  as
 \[   
 I_{r} = I +I_{GH}+  I_{ct}\,, 
 \]
 where $I_{GH}$ denotes the Gibbons-Hawking term. 
For the renormalized on-shell action $I_{r}[\gamma, \psi]$, which is the function of boundary fields $\gamma, \psi$,  
one can introduce the boundary stress tensor, $T^{ij}_{B}$, and the boundary momentum,  $\Pi_{\psi}$, of the field $\psi$, as
\begin{equation} \label{}
\delta I_{r}[\gamma, \psi]  = \frac{1}{16\pi G}\int d^{d} x \sqrt{-\gamma} \Big(T^{ij}_{B} \delta \gamma_{ij} + \Pi_{\psi}\delta \psi\Big)\,,
\end{equation}
where they are finite by construction. 
One may note the identity on the boundary  for the boundary diffeomorphism parameter $\zeta_{i}$
\[   
-2\zeta_{j}\nabla_{i}T^{ij}_{B} + \Pi_{\psi}\Lie_{\zeta}\psi = \nabla_{j}({\cal Z}^{ij}_{B}\zeta_{j})\,,
\]
where ${\cal Z}^{ij}_{B}$ is a certain combination of the appropriate product of $\Pi_{\psi}$ and the field  $\psi$.

Now, let us introduce the boundary current\footnote{This construction is analogous  to the off-shell bulk ADT current construction in the quasi-local formalism for charges~\cite{Hyun:2014nma}.} for the boundary Killing vector $\xi_{B}$, as 
\begin{equation} \label{Bcur}
\sqrt{-\gamma} {\cal J}^{i}_{B}(\xi_{B}) = -\delta \Big(\sqrt{-\gamma}{\bf T}^{~i}_{\!B\, j}\xi^{j}_{\!B}\Big)  + \frac{1}{2}\sqrt{-\gamma}\xi^{i}_{B} \Big(T^{kl}_{B}\delta \gamma_{kl} + \Pi_{\psi}\delta \psi\Big)\,,
\end{equation}
where the improved boundary stress tensor is defined by
\[   
{\bf T}^{ij}_{B} \equiv  T^{ij}_{B} + \frac{1}{2}{\cal Z}^{ij}_{B}\,.
\]
This improved boundary stress tensor is the same as the one given in Ref.~\cite{Papadimitriou:2005ii}.
Here,  the boundary Killing vector is assumed to be unchanged under the variation as $\delta \xi^{j}_{\!B}=0$.  It would be interesting to note that the second term in Eq.~(\ref{Bcur}) becomes the so-called unintegrated trace anomaly $ {\cal A} \equiv 2{\bf T}^{kl}_{B}\gamma_{kl} + \omega\Pi_{\psi}\psi$, when the variation is taken to be the Weyl scaling transformation at the boundary as 
\begin{equation} \label{Weyl}
\delta_{\sigma} \gamma_{ij} = 2 \gamma_{ij}\delta \sigma\,, \qquad \delta_{\sigma} \psi = \omega \psi \delta\sigma\,,
\end{equation}
where $\omega$ denotes the Weyl weight of the field $\psi$.  Using the above current, one can define the infinitesimal conserved charge as
\begin{equation} \label{}
\delta Q_{B} (\xi_{B}) \equiv \frac{1}{8\pi G} \int d^{d-1}x_{i} \sqrt{-\gamma}{\cal J}^{i}_{B}(\xi_{B}) \,.
\end{equation}
As in the bulk, the finite charge expression could be obtained along the integrable one-parameter path in the solution space. 

It is useful to recall that the bulk conserved charge for the Killing vector $\xi$ can be obtained by the codimension two surface integral through the ADT potential $Q^{\mu\nu}_{ADT}$ as
\[   
\delta Q(\xi) \equiv \frac{1}{8\pi G}\int d^{D-2}x_{\mu\nu}\sqrt{-g}\, Q^{\mu\nu}_{ADT}(\xi)\,.
\]
Now, the equivalence of conserved charge expressions from the bulk and the boundary sides can be shown~\cite{Hyun:2014sha} through the equivalence between the ADT potential and the boundary current in the form of 
\begin{equation} \label{BBequiv}
 \sqrt{-g}\, Q^{\eta i}_{ADT}(\xi) \Big|_{\eta\rightarrow \infty} = \sqrt{-\gamma}{\cal J}^{i}_{B}(\xi_{B})\,,
\end{equation}
where the boundary Killing vector $\xi_{B}$ is the boundary limit of the  bulk Killing vector $\xi$.  This result tells us that conserved charges defined in this way should always give us the same  expressions  from the bulk and the boundary. 
In other words, our improved boundary current gives us the expression consistent with the bulk one which is independent of the scheme in the holographic renormalization process.  Furthermore, the  integrability issue to obtain the finite charge expression persists even in the boundary side.  To retain the AdS/CFT correspondence precisely in conserved charges, this equivalence is  the satisfactory feature of our construction.


\bigskip


\end{document}